\documentclass[onecolumn,superscriptaddress]{revtex4}
\usepackage[bookmarks = false,colorlinks = true,linkcolor = blue,urlcolor= black,citecolor = blue]{hyperref}

\usepackage{tcolorbox}
\usepackage{xcolor}
\colorlet{dred}{red!80!black}
\colorlet{dblue}{blue!80!black}
\colorlet{dgreen}{green!45!black}
\colorlet{dyellow}{yellow!60!black}

\newcommand{\ket}[1]{|{#1}\rangle}

\usepackage{graphicx,bm,bbm}
\usepackage{amsmath,amssymb,amsthm}
\usepackage[misc]{ifsym}
\usepackage{appendix}
\usepackage{enumitem}
\usepackage{kotex}
 \usepackage{subcaption}

\begin{document}
\title{Design of Quantum error correcting code for biased error on heavy-hexagon structure }
\author{Younghun Kim}
\email{hpoqh@hanyang.ac.kr}
\author{Jeongsoo Kang}
\email{jskang1202@hanyang.ac.kr}
\author{Younghun Kwon}
\email{yyhkwon@hanyang.ac.kr}
\affiliation{Department of Applied Physics, Center for Bionano Intelligence Education and Research, Hanyang University, Ansan 15588, Republic of Korea}
\begin{abstract}

\indent Surface code is an error-correcting method that can be applied to the implementation of a usable quantum computer. At present, a promising candidate for a usable quantum computer is based on superconductor-specifically transmon. 
Because errors in transmon-based quantum computers appear biasedly as Z type errors, tailored surface and XZZX codes have been developed to deal with the type errors. Even though these surface codes have been suggested for lattice structures, since transmons-based quantum computers, developed by IBM, have a heavy-hexagon structure, it is natural to ask how tailored surface code and XZZX code can be implemented on the heavy-hexagon structure.
In this study, we provide a method for implementing tailored surface code and XZZX code on a heavy-hexagon structure. Even when there is no bias, we obtain $ 0.231779 \%$ as the threshold of the tailored surface code, which is much better than $ 0.210064 \%$ and $ 0.209214 \%$ as the thresholds of the surface code and XZZX code, respectively. Furthermore, we can see that even though a decoder, which is not the best of the syndromes, is used, the thresholds of the tailored surface code and XZZX code increase as the bias of the Z error increases. Finally, we show that in the case of infinite bias, the threshold of the surface code is $  0.264852\%$, but the thresholds of the tailored surface code and XZZX code are $ 0.296157 \% $ and $ 0.328127 \%$ respectively.
\end{abstract}
\maketitle

\section{Introduction}\label{sec:intro}

\indent Quantum computers contain special features such as superposition and entanglement, which are different from those of classical computers. Those features help quantum computers  find prime numbers\cite{ref:shor_algorithm} or perform database searches\cite{ref:grover_algorithm, ref:j.bae} at a significantly higher speed compared to classical computers.\\
 \indent A physical system of a superconductor or ion trap is used to construct a quantum computer. The quantum hardware of the present generation is called Noisy Intermediate-Scale Quantum (NISQ) hardware\cite{ref:nisq}. This is because in the quantum system, errors such as error of interaction between qubits, error of constructing qubits, and error due to the external environment. Therefore, these errors should be handled appropriately. One method to do this is to use a quantum error correcting code(QEC code)\cite{ref:intro,ref:majorana,ref:surface,ref:anyon}.\\
\indent The quantum error-correcting code consists of a series of steps, such as preparation of a logical state, measurement of a syndrome, and the use of a decoder to find a correction operator from measurement values. At present, the most well-known quantum error-correcting code is the  surface code, which constructs a logical qubit using a stabilizer made of four adjacent qubits on the lattice structure\cite{ref:stabilizer,ref:google}.\\
\indent A promising candidate among the NISQ hardware is a quantum computer using a  superconductor. In particular, transmon-based quantum computer is known as a best candidate for constructing realistic quantum computers\cite{ref:crgate,ref:chow,ref:takita,ref:corcoles,ref:krinner}. 
 In NISQ hardware including superconducting quantum computers, errors of the Z operator occur more frequently than those of other types. Because of this fact, there have been quantum error correcting codes dealing with the bias of the Z error\cite{ref:asymmetric,ref:asymmetric2,ref:azad}, which are Tailored Surface code\cite{ref:tailored,ref:tailored_symmetry,ref:tailored_fragile} and XZZX code \cite{ref:xzzx,ref:cat_xzzx}. When the properties of bias and a symmetry of the syndrome are used, those codes provide a better threshold compared to the surface code, even though they have the identical form of assignment in qubits on the lattice structure.\\
\indent  However, the present quantum computers made by IBM, restrict the number of connections  to perform a low error rate operation. Therefore, it uses a heavy-hexagon structure instead of a  lattice structure\cite{ref:heavy-hexagon}. In a heavy-hexagon structure, the maximum number of adjacent qubits is three, instead of four.\\
\indent  Therefore, one can ask how a quantum error-correction code can be constructed in a  heavy-hexagon structure. Constructing a measurement circuit for the syndrome is crucial to implementing a quantum error-correcting code in the heavy-hexagon structure.
Because heavy-hexagon structures have a low-degree connection, the connection restriction  requires the use of flag qubit\cite{ref:fault,ref:bridge,ref:flag,ref:ibm_bridge,ref:surfmap}. Moreover, as explained earlier, the characteristic of existing errors, which implies that errors of the Z operator occur more frequently compared to other types, should be considered. Therefore, in this study, we propose a method to construct Tailored Surface code and XZZX code on a heavy-hexagon structure. In particular, we provide a measurement circuit for the syndrome in the tailored surface code and the XZZX code on the heavy-hexagon structure.\\
 \indent Furthermore, we perform simulations on the proposed measurement circuit for the syndrome to obtain the threshold of the surface code, tailored surface code, and XZZX code not only in the lattice structure but also in the heavy-hexagon structure. First, we can see that when there is no bias in the Z error, the threshold of the tailored surface code is $ 0.231779 \%$, which shows better performance than the thresholds of the surface code and XZZX code being $ 0.210064 \%$ and $ 0.209214 \%$, respectively. In addition, we show that even when using a decoder, which is not most suitable for the syndrome, as the bias of the Z error increases, the thresholds of the tailored surface code and XZZX code increase. Finally, we can see that when the infinite bias is considered, the thresholds of the tailored surface code and XZZX code are $ 0.296157 \% $ and $ 0.328127 \%$, respectively. These values are larger than the threshold of the surface code of $ 0.264852 \%$.
 
\indent This study is organized as follows. First, we explain the assignment of data qubits, flag qubits, and syndrome qubits in the heavy-hexagon structure. Next, we describe the design of the measurement circuit for the syndrome in the proposed qubit structure. Then, we obtain the thresholds of the surface code, tailored surface code, and XZZX code in the lattice and heavy-hexagon structure.

\section{ Result }\label{sec:result}

\subsection{Tailored Surface code \& XZZX code measurement circuit }\label{subsec:circuit}
 To build QEC codes, errors should be detected by measuring syndrome qubits. In this section, we provide a quantum circuit to implement the stabilizer measurement of the tailored surface code and XZZX code in a heavy-hexagon structure. The tailored surface code and XZZX code have the same qubit structure despite the different stabilizers compared to the surface code.\\
\indent In the following subsection, we first provide the positions of the data, flag, and syndrome qubits in a heavy-hexagon structure. Second, we explain the quantum circuit of stabilizer measurement to measure the syndrome using data qubits, flag qubits, and syndrome qubits. 

\subsubsection{ Heavy-hexagon architecture  }\label{subsubsec:archi}

 Usually, QEC codes based on surface code need to check the parity of the four data qubits. The lattice structure is designed such that each qubit is connected to four different neighboring qubits. Meanwhile, in a heavy-hexagon structure, qubits have less connectivity than in lattice structures. In  other words, each qubit of the heavy-hexagon structure cannot be entangled with four qubits directly. To solve the issue, we use more auxiliary qubits so-called flag qubits, so that some qubits can interact indirectly with four different qubits\cite{ref:surfmap}. Therefore, three kinds of qubits, such as data qubits to build logical qubit state, syndrome qubits to detect errors, and flag qubits to connect data qubits and the syndrome qubits are required to implement  a surface code in a heavy-hexagon structure. Fig.1 shows the arrangement of data qubits, flag qubits, and syndrome qubits for an error-correcting code of [[13,1,3]] in the Ithaca system with a heavy-hexagon structure of 65 superconducting qubits of IBM\cite{ref:demonstration}. Four distinguishable data qubits are linked to the syndrome qubits along the flag qubits at each stabilizer patches.

\begin{figure*}[ht]
    \centering
    \includegraphics*[width=0.8\textwidth]{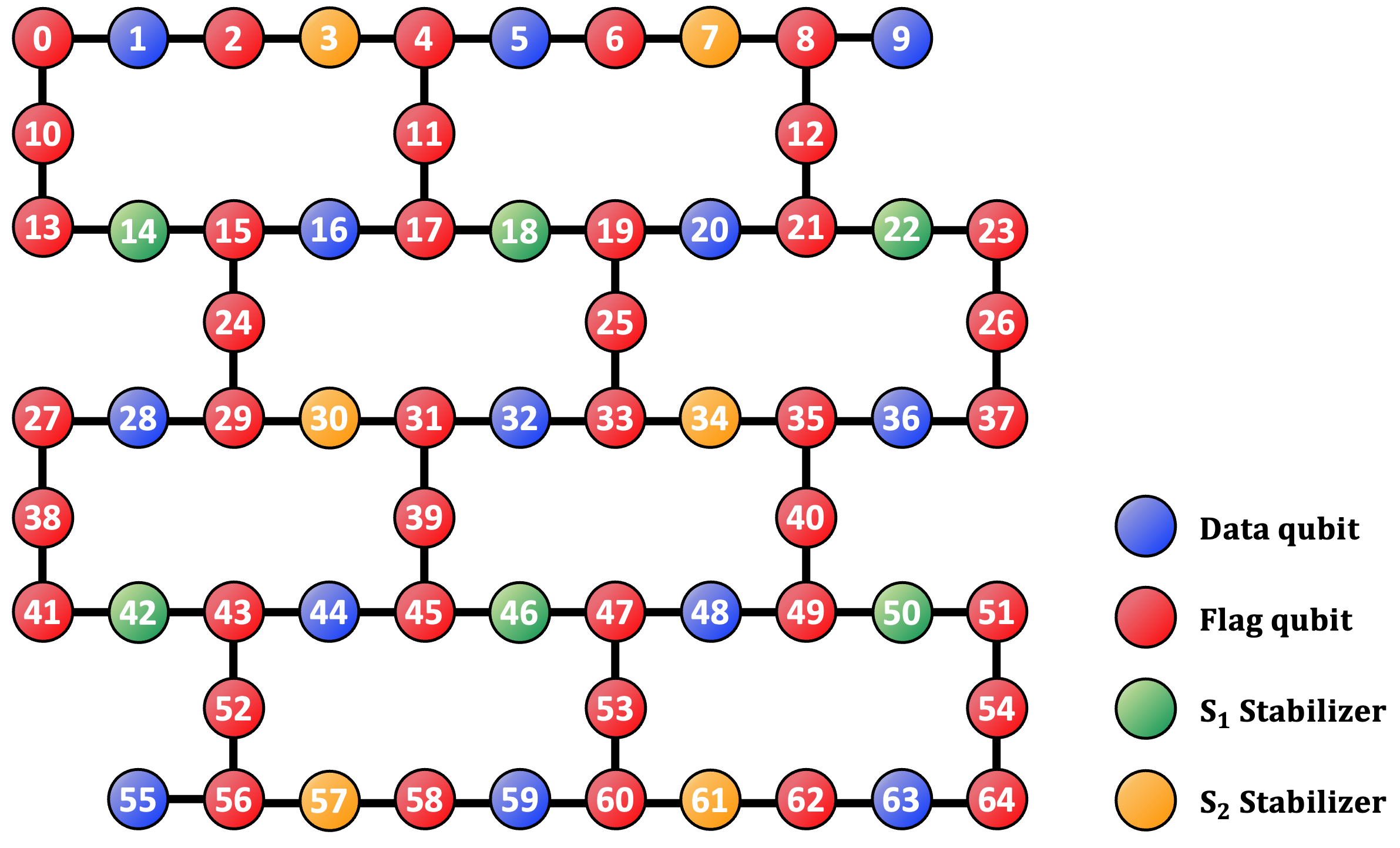}
    \caption{ [[13,1,3]] Surface code in the heavy-hexagon structure. The data qubits(flag qubits) are represented by   blue dots(red dots). Meanwhile, the syndrome qubits are denoted by orange and green dots.}
    \label{fig:1}
\end{figure*}

In Fig.1, the blue and red dots denote the data and flag qubits, respectively. $S_1$ and $S_2$ indicate the stabilizers. The black edges indicate that two lined-up qubits can be entangled. The implementation consists of 13 data qubits, 36 flag qubits, and 12 syndrome qubits in the Ithaca system. Distance 3 surface-code based QEC codes, which can correct at least a single error, use the same number of the data qubits and syndrome qubits except for the flag qubits in the lattice structure. The tailored surface code, XZZX code, and surface code have identical qubits arrangements regardless of the type of stabilizer, but different stabilizer measurement circuits. The logical qubit state $\ket{\psi}_L$ is defined as a data qubit state that has $S_1$ and $S_2$ stabilizers whose eigenvalues are equal to $+1$. An eigenvalue of $-1$ is occured when there are some errors that anti-commute with the corresponding stabilizer. 

\subsubsection{ Measurement circuit  of Tailored Surface code }\label{subsubsec:tailored}

 The tailored surface code uses a Y stabilizer instead of a Z stabilizer. The Y stabilizer checks the parity of the X and Z errors simultaneously, and the X stabilizer checks the parity of the Z errors. 

\begin{equation}
     \hat{X}_a \hat{X}_b \hat{X}_c \hat{X}_d , \hat{Y}_a \hat{Y}_b \hat{Y}_c \hat{Y}_d
\end{equation}

The stabilizer for the four data qubits a,b,c, and d can be described as above. The stabilizers are used to check the parity of the error from the ancilla qubits in the stabilizer measurement circuit. The Y stabilizers provide the result of the parity of the X or Z error. Two stabilizer types can detect and correct X and Z errors. In particular, the parity information of the Z errors from both stabilizers can improve the error-correcting performance when the Z errors are dominant\cite{ref:tailored_symmetry}. Here $ S_1 $ ($ S_2 $) denotes the group of Y stabilizers(X stabilizers). Each stabilizer consists of at least three  Pauli operators. 

\begin{align}
    S_1 &= \{ \hat{Y}_1 \hat{Y}_{16} \hat{Y}_{28} ,  \hat{Y}_5 \hat{Y}_{16} \hat{Y}_{20} \hat{Y}_{32} , \hat{Y}_{9} \hat{Y}_{20} \hat{Y}_{36}, \hat{Y}_{28} \hat{Y}_{44} \hat{Y}_{55},  \hat{Y}_{32} \hat{Y}_{44} \hat{Y}_{48} \hat{Y}_{59} ,  \hat{Y}_{36} \hat{Y}_{48} \hat{Y}_{63}\}\\
    S_2 &= \{ \hat{X}_1 \hat{X}_5 \hat{X}_{16} ,  \hat{X}_5 \hat{X}_9 \hat{X}_{20} , \hat{X}_{16} \hat{X}_{28} \hat{X}_{32} \hat{X}_{44} , \hat{X}_{20} \hat{X}_{32} \hat{X}_{36} \hat{X}_{48}, \hat{X}_{44} \hat{X}_{55} \hat{X}_{59} ,  \hat{X}_{48} \hat{X}_{59} \hat{X}_{63} \}
\end{align}

\begin{figure*}[ht]
    \centering
    \includegraphics*[width=0.8\textwidth]{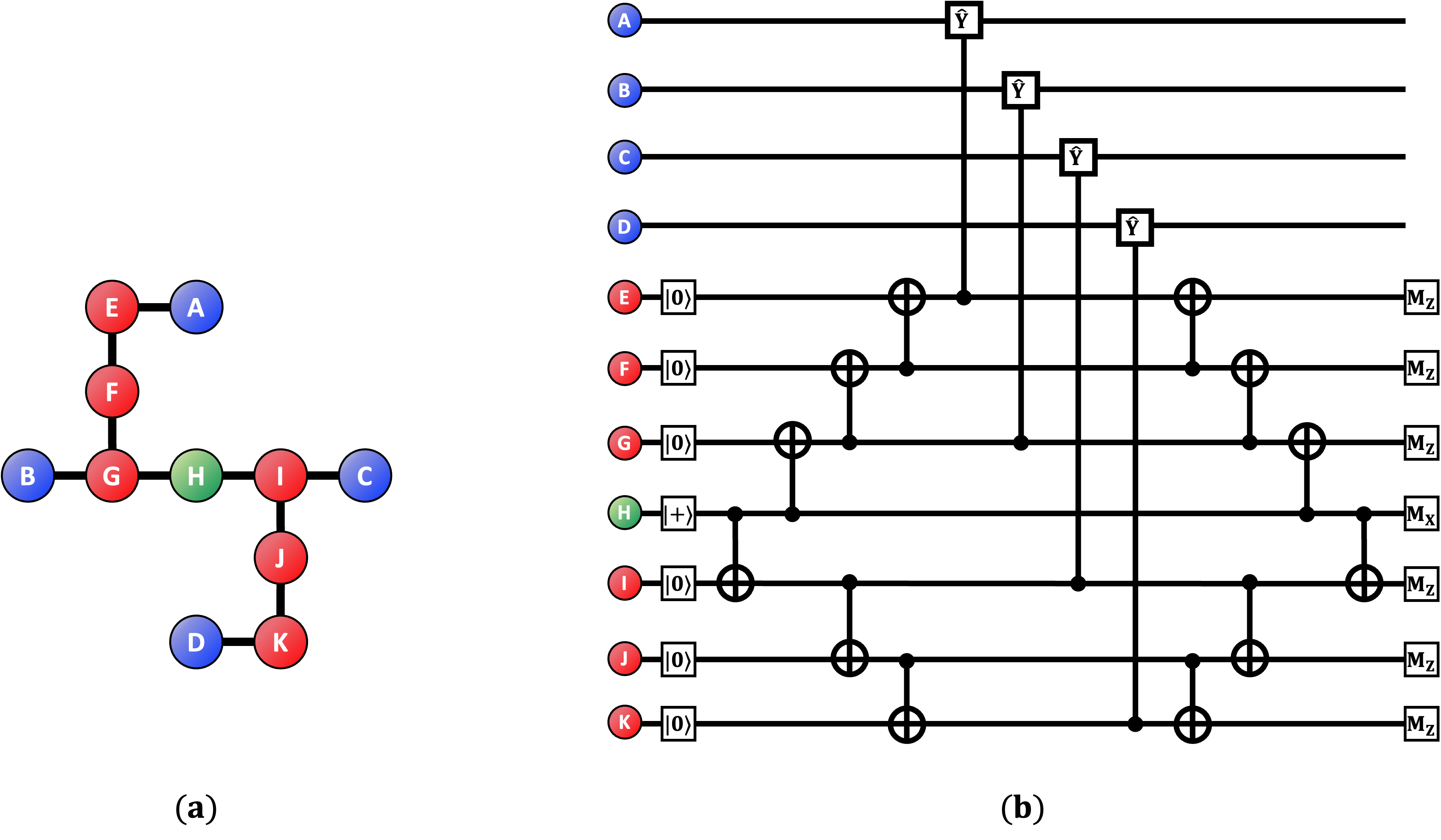}
    \caption{ Y Stabilizer using six flag qubits. (a) Connection among qubits.  (b) Measurement circuit.
    The blue dots denote data qubits, and the red dots denote flag qubits. H is a syndrome qubit. For a measurement of the Y syndrome, the quantum states of the flag qubits(syndrome qubits) are prepared as $\ket{0}$($\ket{+}$).}
    \label{fig:2}
\end{figure*}

 Fig.2 shows the Y stabilizer measurement circuit, where four data qubits, six flag qubits, and one Y syndrome qubit are displayed as blue, red, and green dots, respectively. In the Y stabilizer measurement circuit, the interaction between the data qubit and flag qubit is performed through a controlled Y operator, which is different from the CNOT operator used in the Z and X stabilizer measurement circuits. If the Y syndrome qubit is measured in the X basis and the result of measurement of Y stabilizer becomes $\ket{+}$($\ket{-}$),  the eigenvalue is +1(-1). When the controlled Y is used for the Y stabilizer measurement, the qubit error to the data qubit or flag qubit produces various results in the Y stabilizer measurement. 
  
 Fig.3(a) shows the case in which data qubit has an X error.  Fig3(b) shows how the measurement result is changed in the Y stabilizer measurement circuit when there is an X error in the data qubit. The X error in the data qubit causes the measurement result to flip and $\ket{-}$ is measured. If an X error occurs additionally in different data qubit, the measurement result of the syndrome flips twice and $\ket{+}$ is measured. Therefore, if the X errors may be occurred even(odd) times, we get $\ket{+}$($\ket{-}$) in the Y stabilizer measurement.\\
\indent It can be observed that when there is an X error in the data qubit the parity to the error can be checked by the measurement result of the Y syndrome; however, there is no change in the measurement result for the flag qubit. It should be noted that the error in the data qubit causes a propagation of the Z error after an interaction between the data qubit and flag qubit. However, $\ket{0}$ is robust with respect to the Z error. Therefore, when Y stabilizer measurement circuit is prepared, the initial state of flag qubits needs to be $\ket{0}$ in order that an error in a data qubit should not affect the measurement of the flag qubits. 

\begin{figure*}[ht]
    \centering
    \includegraphics*[width=0.8\textwidth]{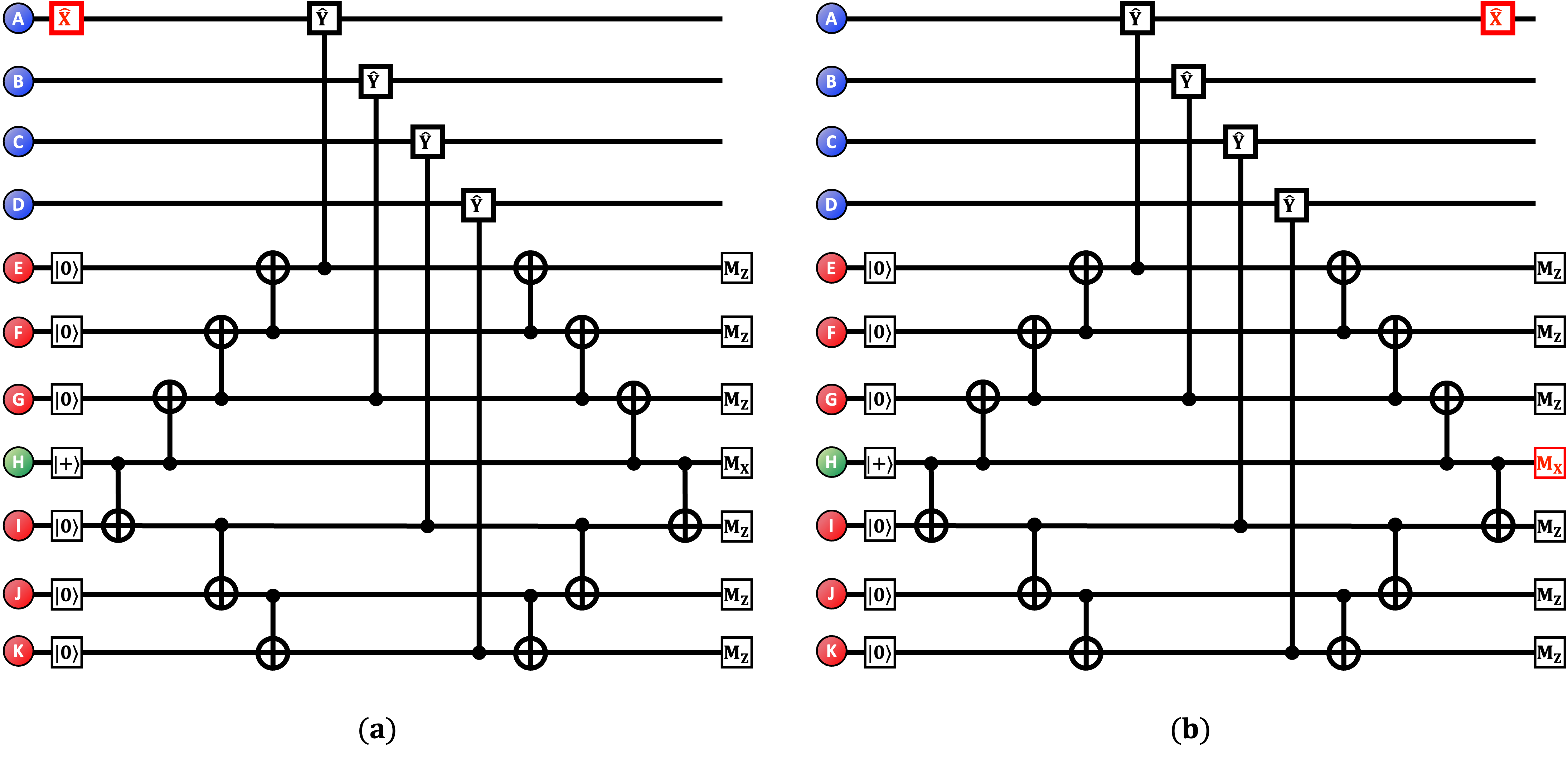}
    \caption{(a) Case where an X error occurs in Y stabilizer measurement circuit (b) Change in the measurement when an X error occurs in Y stabilizer measurement circuit}
    \label{fig:3}
\end{figure*}

 Fig.4 shows the case in which an Z error occurs in the data qubit. When an Z error occurs in the data qubit, Fig.4(a) shows how the measurement circuit of the Y stabilizer is performed and Fig.4(b) shows how the error affects the measurement circuit of the Y stabilizer. The blue color denotes the flip of the measurement result in the Y syndrome qubit, which changes from $\ket{+}$ to $\ket{-}$. Both the X and Z errors in the data qubit affect the measurement results of the Y syndrome qubit. Therefore, the Y error in the data qubit causes it to flip twice and does not affect the measurement result of the Y syndrome qubit. 

\begin{figure*}[ht]
    \centering
    \includegraphics*[width=0.8\textwidth]{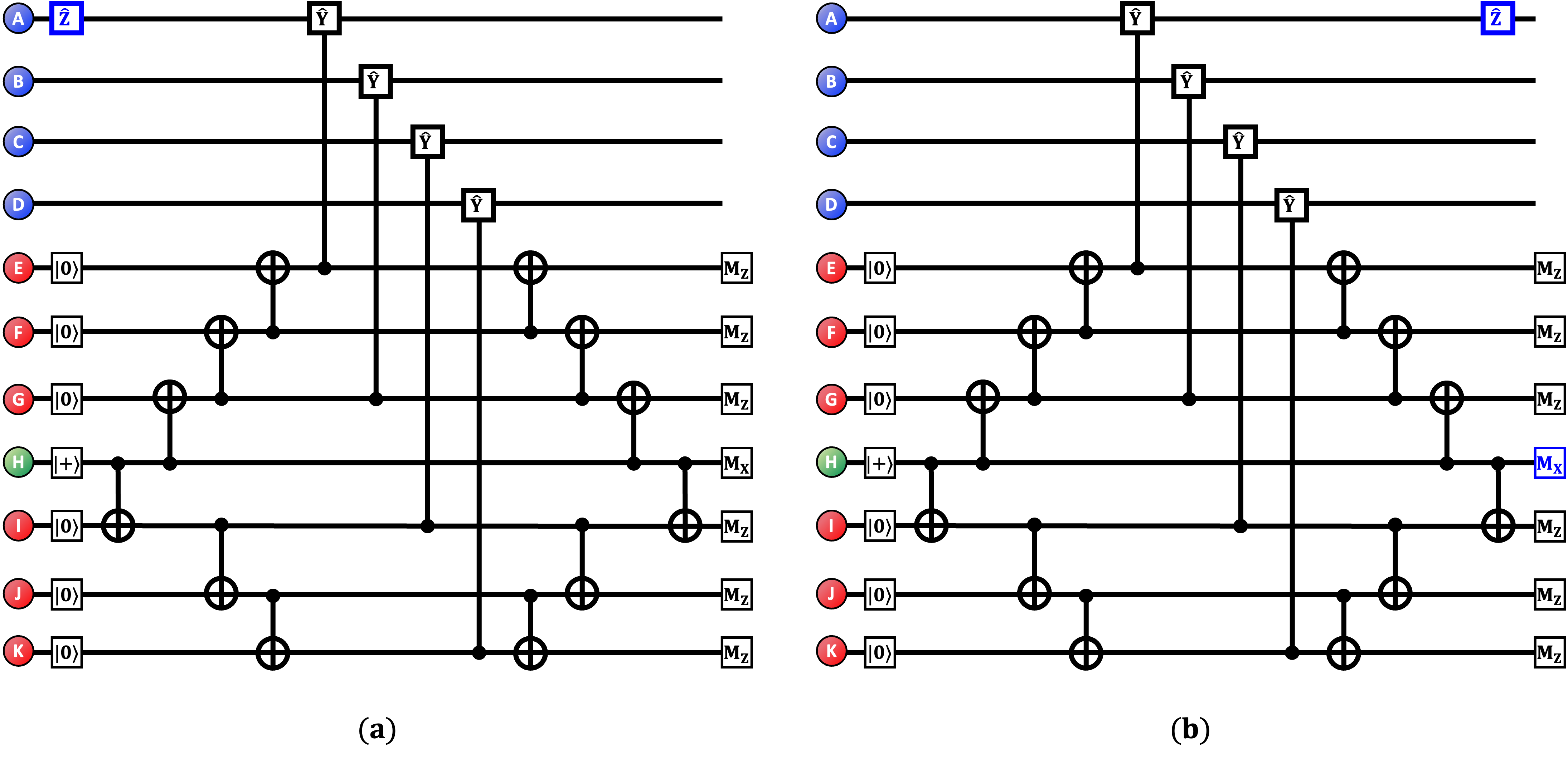}
    \caption{ (a) Case where an Z error of a data qubit occurs in Y stabilizer measurement circuit (b) Change in the measurement when an Z error of a data qubit occurs in Y stabilizer measurement circuit}
    \label{fig:4}
\end{figure*}

\begin{figure*}[ht]
    \centering
    \includegraphics*[width=0.8\textwidth]{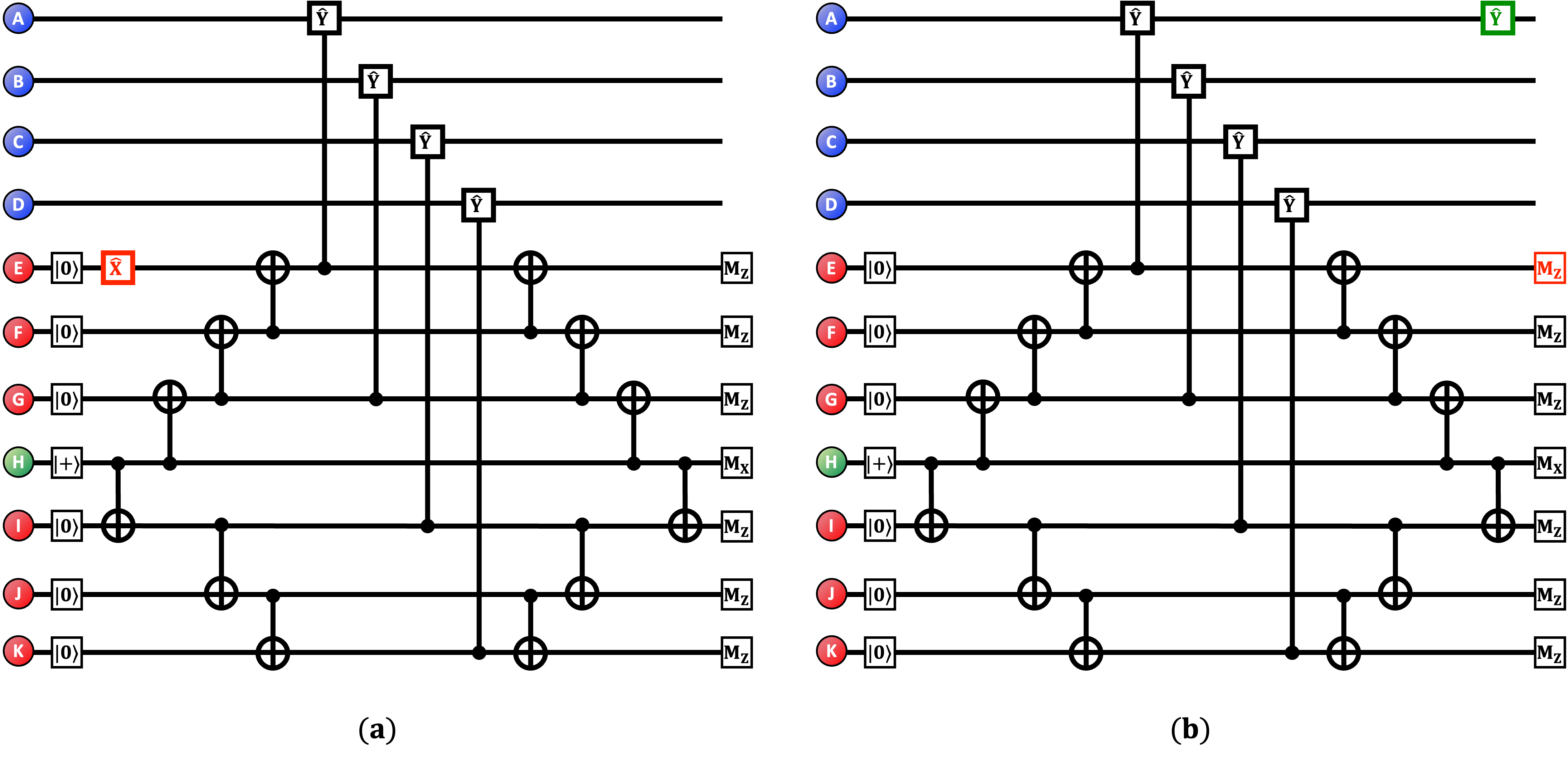}
    \caption{(a) Case where an X error of a flag qubit occurs in Y stabilizer measurement circuit (b) Changes in the data qubit and the measurement when an X error of a flag qubit occurs in Y stabilizer measurement circuit}
    \label{fig:5}
\end{figure*}

Not only the data qubit but also the flag qubit may have an error in the Y stabilizer measurement circuit. The error in the flag qubit may be propagated to the data qubit. In order to detect the error propagated from the flag qubit to the data qubit, every initial state of the flag qubit needs to be set as $\ket{0}$. The CNOT is constructed by choosing the flag qubit adjacent to the syndrome qubit as the target qubit and the syndrome qubit as the control qubit. 

 Fig.5 shows the case in which an X error occurs in the flag qubit. The X error of the flag qubit causes a Y error propagation to the data qubit. Error propagation to the data qubit by the error of the flag qubit can be checked through the measurement result of the flag qubit. That is, the Y stabilizer measurement circuit enables checking not only the parity to X or Z error of the data qubit but also the error propagation to the data qubit by the error of the flag qubit. The tailored surface code also requires X stabilizer measurement circuit, as shown in Fig.6(a). Similar to the Y stabilizer measurement circuit, four data qubits(six flag qubits) are denoted by blue dots(red dots) and one X syndrome qubit is displayed by orange dot. In the X stabilizer measurement circuit, the interaction between the data qubit and flag qubit is performed by the CNOT operation, where the flag qubit is used as the control qubit and the corresponding the data qubit is used as the target qubit. 

\begin{figure*}[ht]
    \centering
    \includegraphics*[width=0.8\textwidth]{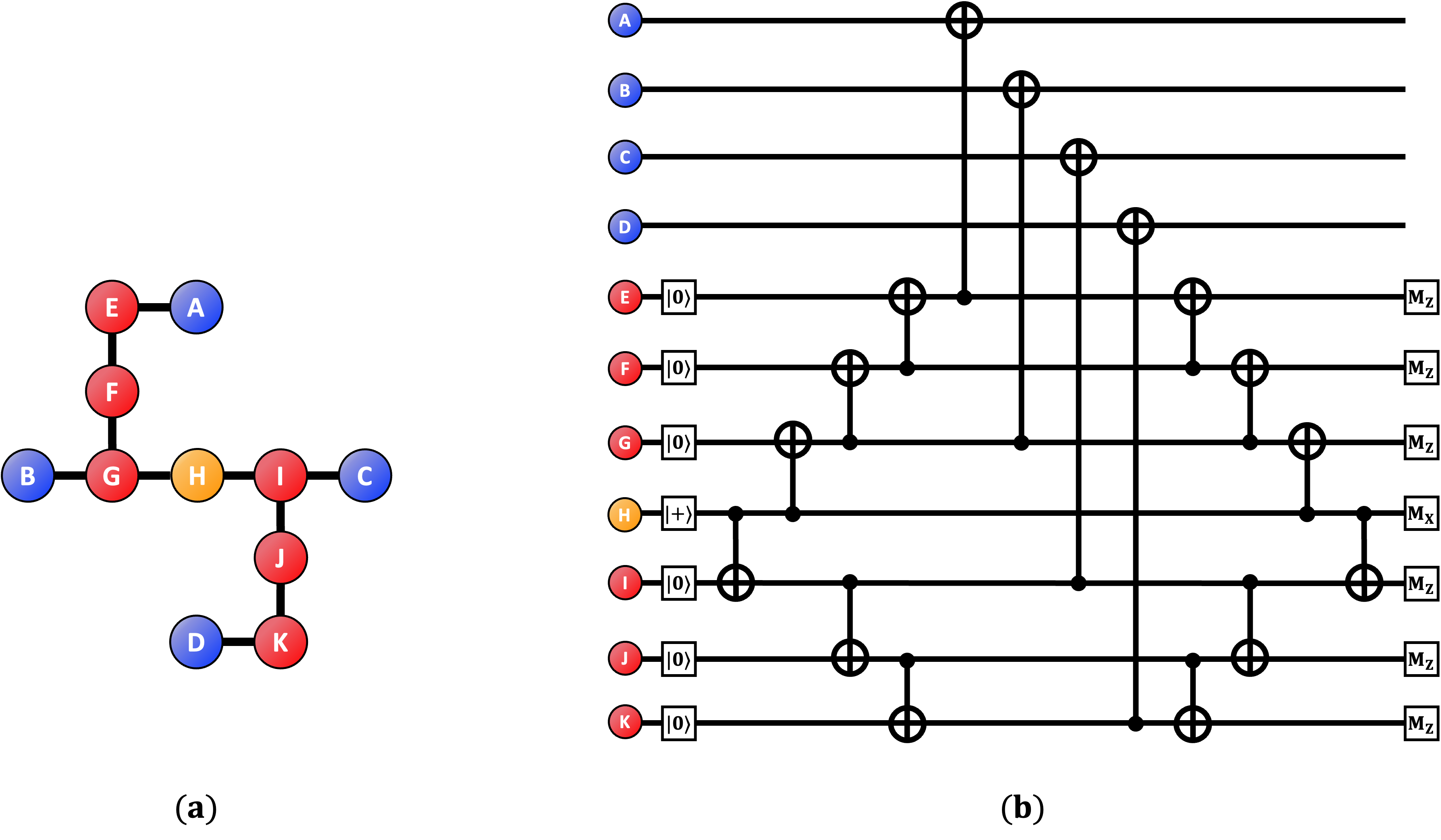}
    \caption{ X Stabilizer using six flag qubits. (a) Connection among qubits.  (b) Measurement circuit. The blue dots denote data qubits, and the red dots denote flag qubits. H is a syndrome qubit. For a measurement of the X syndrome, the quantum states of flag qubits(syndrome qubits) are prepared as $\ket{0}$($\ket{+}$).}
    \label{fig:6}
\end{figure*}

 Fig.6(b) shows how the parity to check the Z error of the data qubit can be obtained in the X stabilizer measurement circuit. The initialization of each flag qubit is prepared as $\ket{0}$ and the initialization of the X syndrome qubit as $\ket{+}$. The difference between the X and Y stabilizers exists when there is an X error in the data qubit and it does not affect the X stabilizer's syndrome result but the Y stabilizer's syndrome result. This is because of the difference between the CNOT and Controlled-Y operators when constructing the entanglement between the flag qubits and the data qubits.

\subsubsection{ Measurement Circuit of XZZX code}\label{subsubsec:xzzx}

Every stabilizer of XZZX Code has only one type. The XZZX stabilizer consists of operators to four data qubits and the XZZX stabilizer corresponding to a,b,c, and d qubits is given as follows:

\begin{equation}
     \hat{X}_a \hat{Z}_b \hat{Z}_c \hat{X}_d 
\end{equation}

Fig.7 shows the measurement circuit of the XZZX stabilizer. And the following $ S_1 $ and $ S_2 $ are the group of XZZX stabilizers.

\begin{align}
    S_1 &= \{ \hat{X}_1 \hat{Z}_{16} \hat{X}_{28} ,  \hat{X}_5 \hat{Z}_{16} \hat{Z}_{20} \hat{X}_{32} , \hat{X}_{9} \hat{Z}_{20} \hat{X}_{36}, \hat{X}_{28} \hat{Z}_{44} \hat{X}_{55},  \hat{X}_{32} \hat{Z}_{44} \hat{Z}_{48} \hat{X}_{59} ,  \hat{X}_{36} \hat{Z}_{48} \hat{X}_{63}\}\\
    S_2 &= \{ \hat{Z}_1 \hat{Z}_5 \hat{X}_{16} ,  \hat{Z}_5 \hat{Z}_9 \hat{X}_{20} , \hat{X}_{16} \hat{Z}_{28} \hat{Z}_{32} \hat{X}_{44} , \hat{X}_{20} \hat{Z}_{32} \hat{Z}_{36} \hat{X}_{48}, \hat{X}_{44} \hat{Z}_{55} \hat{Z}_{59} ,  \hat{X}_{48} \hat{Z}_{59} \hat{Z}_{63} \}
\end{align}

The measurement circuit of the XZZX stabilizer is composed of four data qubits, six flag qubits, and one syndrome qubit. The interaction between the data qubit and flag qubit is important in the measurement circuit of the XZZX stabilizer, because the measurement result of the flag qubit is used to detect error propagation. It should be noted that Hadamard operator is required  to check the error propagation to the data qubit by the error of the flag qubit. The Hadamard operator can transform a bit-flip error into a phase-filp error. The circuit is designed such that all flag qubits are initialized as $\ket{0}$, and Hadamard gates are applied such that data errors do not disturb the flag qubit states. Therefore, one can detect the data qubit errors with stabilizer measurement results identically as in the case of lattice-structure-based XZZX code. 

\begin{figure*}[ht]
    \centering
    \includegraphics*[width=0.8\textwidth]{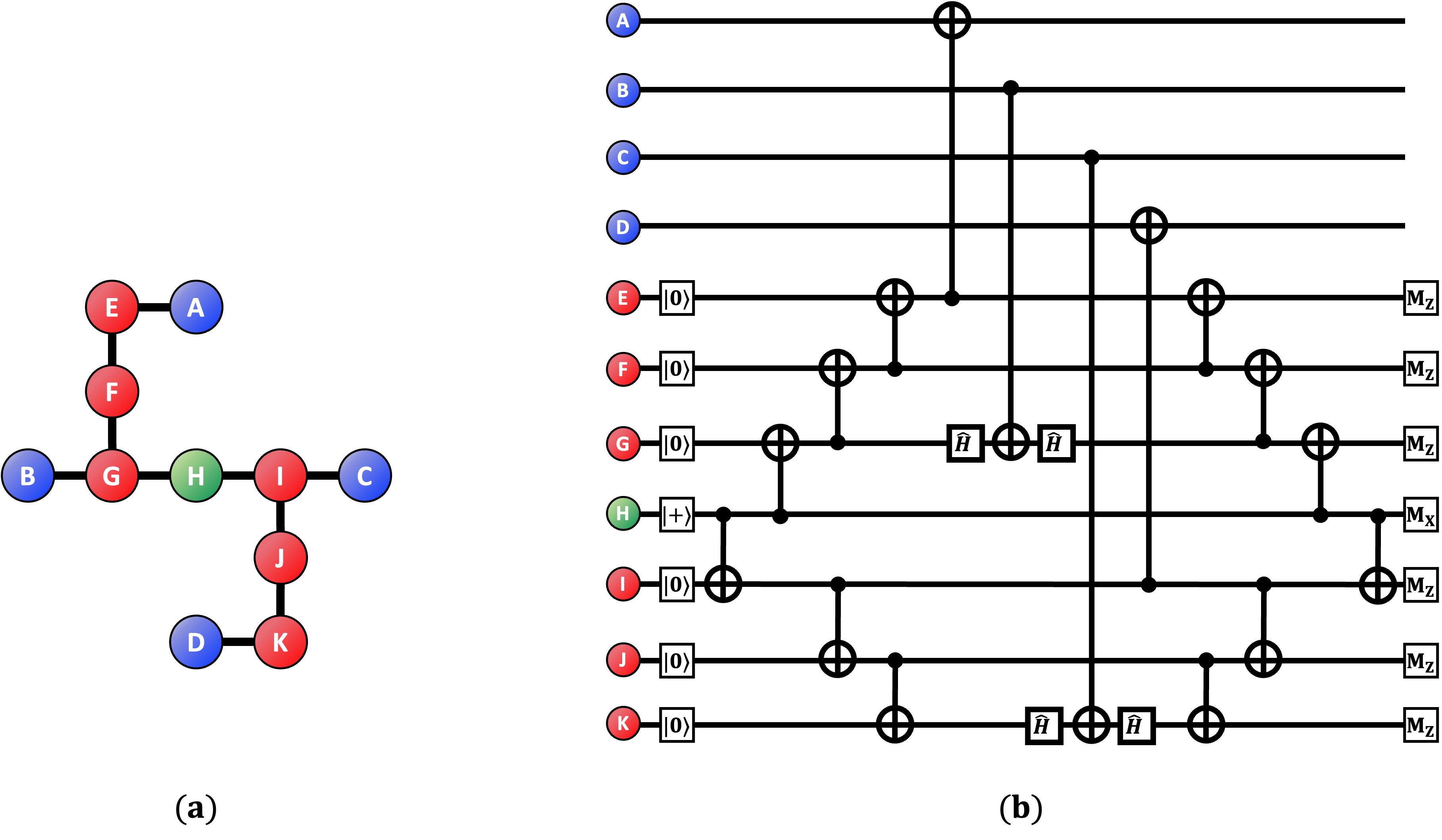}
    \caption{XZZX Stabilizer using six flag qubits. (a) Connection among qubits.  (b) Measurement circuit. The blue dots denote data qubits, and the red dots denote flag qubits. For a measurement of the XZZX syndrome, the quantum states of the flag qubits(syndrome qubits) are prepared as $\ket{0}$($\ket{+}$). The interaction between the data qubit and the flag qubit is established by CNOT and H operators.}
    \label{fig:7}
\end{figure*}

The surface code, tailored surface code and XZZX code have two stabilizer groups. We measure two stabilizer group syndrome simultaneously in the lattice structure. However, we check them individually in the heavy-hexagon structure, because the two stabilizer group measurement circuits share partial flag qubits. Therefore, we do not need to consider the hook error in a heavy-hexagon structure.

\subsection{ Physical Qubit Error Model }\label{subsec:error}

\subsubsection{ Depolarization Error Model }\label{subsubsec:dep}

 The X,Y and Z errors of a single qubit occur equally in the depolarization error model. Therefore, when the probability of error in a qubit is p and $p_X , p_Y , p_Z$ denote the error probabilities of the X,Y, and Z errors respectively, the following relation holds:

\begin{equation}
     \frac{p}{3} = p_X = p_Y = p_Z
\end{equation}

The depolarization error model is an error channel on the qubit, where the probabilities of the X,Y, and Z errors are identical.   

\subsubsection{ Biased Dephasing Error Model }\label{subsubsec:bias}
 Real quantum computer hardware, such as a transmon-based quantum computer, is susceptible to the Z errors, which implies that the Z errors are more frequent than the X or Y errors. This situation can be described using a biased dephasing error model. The biased dephasing error model is expressed by the degree of bias $\eta $ which shows the different error probabilities in $\ p_X , p_Y$, and $p_Z$. These satisfy the following relationship:

\begin{equation}
\begin{array}{cc}
     & \frac{p}{2(\eta+1)} = p_X = p_Y \\
     \\
     & \frac{p\eta}{\eta+1} = p_Z
\end{array}
\end{equation}

Therefore, the depolarization error model corresponds to the case of $\ \eta = 0.5$ in the biased dephasing error model. Therefore, if $\eta = \infty $ then only the Z error will occur. 

\begin{equation}
\begin{array}{cc}
     p_X + p_Y + p_Z = p \\
\end{array}
\end{equation}

In the error channel of Pauli operators, the sum of all types of errors is $p$. The details of the biased dephasing error model are as follows:

\begin{enumerate}
    \item In a single gate, the qubit on which the gate acts, passes through an independent error channel in the error probabilities of $ p_X , p_Y$,  and $p_Z $ at $ \{X,Y,Z\} $ operators. \\
     \item In a CNOT gate, the two qubits, which the gate acts on, pass independently and uniformly through the error channel of $ \{I,X,Y,Z\}^{\otimes{2}} \backslash \{ I \otimes I \} $ in the error probability of $p$.
     \item The qubits are prepared as the state of $ \hat{X} \ket{0} = \ket{1} $ in a probability of $ p_X + p_Y $. 
     \item  A flip error occurs in every measurement result on a single qubit, with a probability of $ p_X + p_Y $.  
     \item  In a stabilizer measurement circuit, the data qubits pass independently through an error channel in the error probabilities of $ p_X , p_Y$,  and $p_Z $ at $ \{X,Y,Z\} $ operators..
\end{enumerate}

Here the error models of the controlled-Y operator and CNOT operators are treated in the same manner. In addition, it is assumed that every error has an independent and identical distribution.

\subsection{ Simulation results }\label{subsec:sim}

In this section we evaluate the thresholds of the tailored surface code, XZZX code, and surface code in terms of the  biased dephasing error model, including the depolarization error model. In particular, when the bias is in $\eta = \{ 0.5, 1, 10, 100, 1000,\infty \} $, we obtain the thresholds of the tailored surface code, XZZX code, and surface code on the lattice and heavy-hexagon structure. To evaluate the thresholds, we use the measurement circuit for the syndrome described in the previous section and the stim code\cite{ref:stim} package. In addition, to obtain the correction operator, we use PyMatching\cite{ref:pymatching}, considering the X and Y errors independently in the tailored surface code and the X and  Z errors independently in the XZZX code and surface code. In the decoder, we do not consider the correlation between the errors. In the cases of the surface code and XZZX code, a logical X error and a logical Z error exist. In the tailored surface code, the logical X and Y errors exist. We obtain the logical failure by evaluating the similarity between the logical state to initialized state of the data qubit and the final state applied by the correction operator. For instance, in the surface code, the logical X error rate is evaluated by checking whether the corrected state is the same as $\ket{0}_L$ when the initialized state is prepared as $\ket{0}_L$. Meanwhile, the logical Z error rate is evaluated by checking whether the corrected state is the same as $\ket{+}_L$ when the initialized state is prepared as $\ket{+}_L$. In QEC code, to consider every error of two logical operators, $E_{total}$ is obtained by the logical error rates $E_{X}$ and $E_{Z}$ as follows\cite{ref:logic_error}:

\begin{equation}
\begin{array}{cc}
     E_{total} = 1-(1- E_{X} ) \times (1-E_{Z} ) \\
\end{array}
\end{equation}

The following figures show the thresholds in terms of error probabilities $p$ and $\eta $. Here, we perform simulations using at least 1200000 samples and obtain results for $d = \{3,5,7,9,11\}$. Here, to correct the measurement, we perform the error correction after measuring the syndrome in the repetition of d round.

\begin{figure*}[ht]
   \centering
   \includegraphics[width=0.8\textwidth]{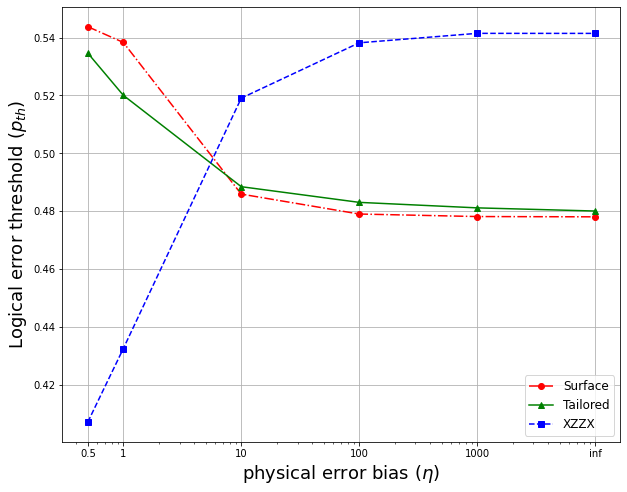}
   \caption{The thresholds of the surface code, tailored surface code, and XZZX code in the lattice structure are displayed according to the bias. The red circle displays the thresholds of the surface code, and the green triangle displays the thresholds of the tailored surface code. The blue rectangle displays the thresholds of the XZZX code.}
   \label{fig:8}
   
\end{figure*}    

Fig.8 shows the behavior of the threshold in all QEC codes on a lattice structure. The threshold of the surface code is $ 0.543786 \% $ at $ \eta =0.5 $, which implies no bias. As $ \eta $ increases to $ \eta = \infty $, the threshold of the surface code becomes $  0.478029\% $. In addition, the threshold of the tailored surface code becomes $ 0.534635 \%$ at $ \eta =0.5 $ and diminishes to $ 0.480034 \% $ at $ \eta = \infty $. 
In the XZZX code, the threshold at $ \eta =0.5 $ becomes $ 0.407061 \% $, which is the lowest value. In addition, as the bias increases, the threshold also increases. In the case of $ \eta = \infty $, the threshold becomes the largest value of $ 0.541467 \% $, compared to the thresholds of other codes.

\begin{figure*}[ht]
   \centering
    \includegraphics[width=0.8\textwidth]{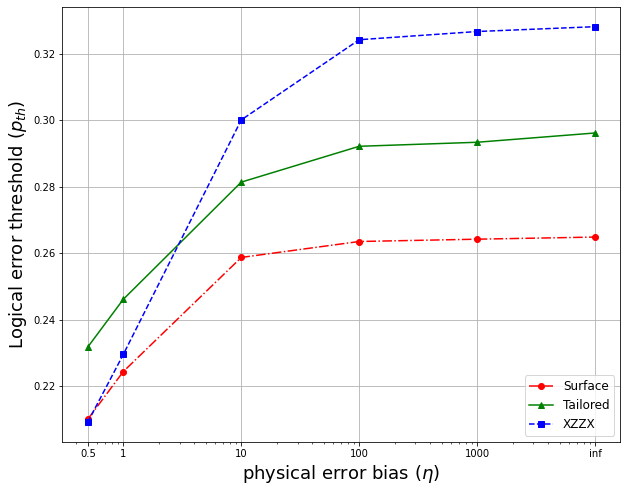}
    \caption{Thresholds of the surface code, and tailored surface code, XZZX code in the heavy-hexagon structure, according to the bias. The red circle displays the thresholds of the surface code, and the green triangle displays the thresholds of the tailored surface code. The blue rectangle displays the thresholds of the XZZX code.}
    \label{fig:9}
\end{figure*}

Finally, the thresholds of the QEC codes for the logical error are investigated on the heavy-hexagon structure. Fig.9 shows the results. First, in the case of the tailored surface code, the threshold for the logical error becomes $ 0.231779 \% $ at $ \eta =0.5 $. Second, in the case of the surface code code, the threshold for the logical error becomes $ 0.210064 \% $ at $ \eta =0.5 $. In addition, the threshold of the XZZX code for the logical error has the lowest value of $ 0.209214 \% $ at $ \eta =0.5 $, compared to the thresholds of the tailored surface code and surface code. When there is no bias, the threshold of the tailored surface code becomes the highest value. When $ \eta = \infty $, the threshold of the XZZX code to the logical error becomes $ 0.328127 \% $. The threshold of the tailored surface code in the lattice structure decreases as the bias increases, but the threshold of the tailored surface code in the heavy-hexagon structure increases. Threshold of the tailored surface code becomes $ 0.296157 \% $ at $ \eta = \infty $ which is higher than the threshold of the surface code $ 0.264852 \%$ at $ \eta = \infty $. Therefore, in the heavy-hexagon structure,  the threshold of the tailored surface code is higher than that of the surface code in all regions of bias. The tailored surface code has a higher threshold value than the XZZX code when the bias is low. In the heavy-hexagon structure, the thresholds of the QEC codes do not change rapidly after the value of $\eta = 100$. It should be noted that in the heavy hexagon structure the threshold of the XZZX code provides a larger value than those of the other error-correcting codes in the range of $\eta$ which is larger than $\eta = 10$.\\
\indent Fig.10 shows the thresholds of the surface code, tailored surface code, and XZZX code in the heavy-hexagon structure when the bias is 100. The thresholds of each code are $ 0.263515 \%$, $ 0.292152 \%$, and $ 0.324207 \%$ respectively.  

\begin{figure*}[ht]
   \centering
     \begin{subfigure}[b]{0.45\textwidth}
         \centering
         \includegraphics[width=\textwidth]{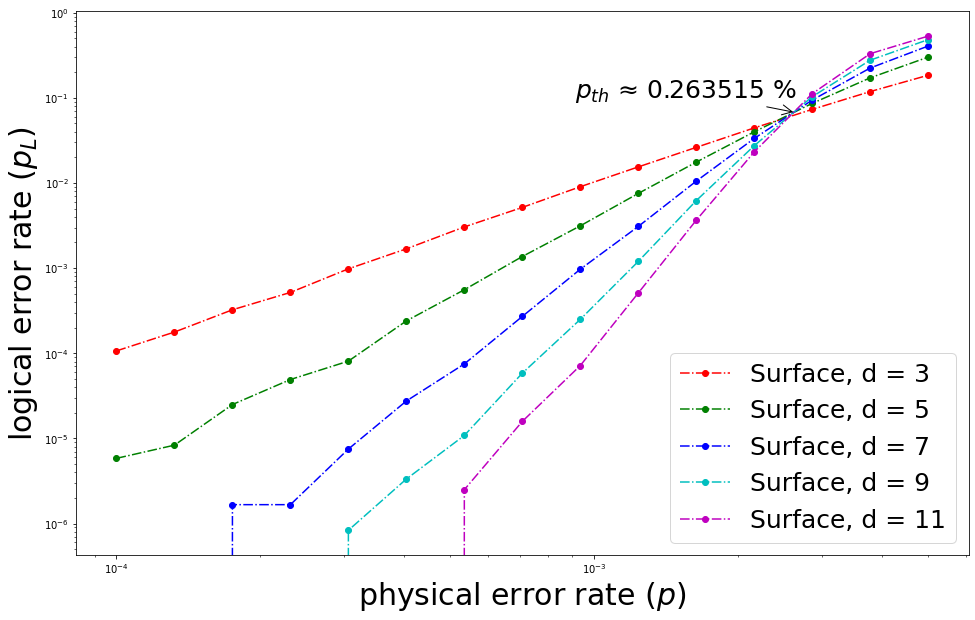}
         \caption{Surface code}
         \end{subfigure}
     \hfill
     \begin{subfigure}[b]{0.45\textwidth}
         \centering
         \includegraphics[width=\textwidth]{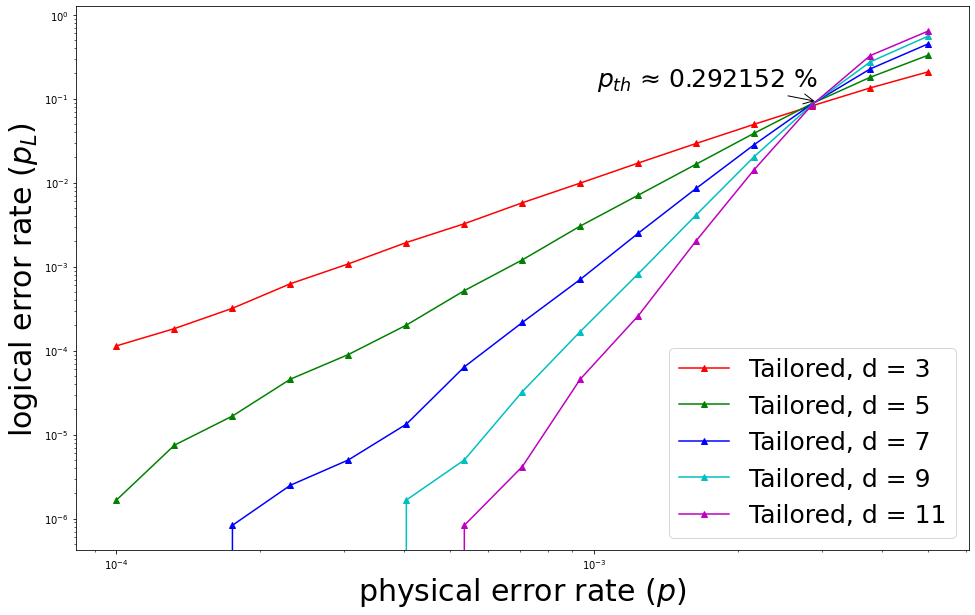}
         \caption{Tailored Surface code}
         \end{subfigure}
     \hfill
     \begin{subfigure}[b]{0.45\textwidth}
         \centering
         \includegraphics[width=\textwidth]{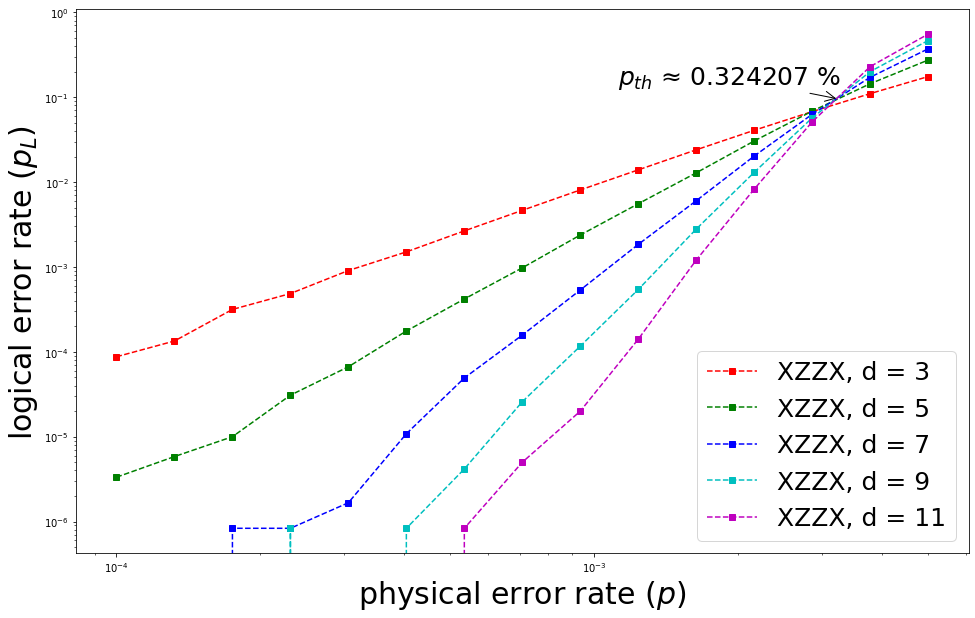}
         \caption{XZZX code}
         \end{subfigure}
    \caption{(a) Thresholds of surface code in the heavy-hexagon structure (b) Thresholds of tailored surface code in the heavy-hexagon structure (c) Thresholds of XZZX Code in the heavy-hexagon structure.
     The thresholds of surface code, tailored surface code, and XZZX code in the heavy-hexagon structure are displayed when the bias is 100.}
    \label{fig:10}

\end{figure*}

\section{Conclusion}
  In this work, we provided the design of quantum error correcting codes in the heavy-hexagon structure which is used in the hardware of IBM quantum computers. First, we showed the method to assigning data qubits, syndrome qubits, and flag qubits in the heavy-hexagon structure to construct a measurement circuit for the syndrome of quantum error correcting code. Second, we built a measurement circuit for the syndrome by considering the entanglement among the three types of qubits. In the tailored surface code, we used the Controlled-Y operator to create a measurement circuit for the the Y syndrome. In the XZZX code, we used the Hadamard operator to detect the data qubit errors. \\
\indent Furthermore, we compared the thresholds of quantum error-correcting codes designed in the heavy-hexagon structure with those of quantum error correcting codes in the lattice structure. Our results showed that the flag qubits and syndrome qubits might affect the performance of quantum error correcting codes in a heavy-hexagon structure. Because the heavy-hexagon structure requires a restriction on the connection, the flag qubits to overcome the restriction are indispensable. It should be noted that because the number of the flag qubits in the heavy-hexagon structure is more than half of the total number of qubits, the errors of the flag qubits can have a sensitive effect on the performance of quantum error correcting codes. Therefore, our results imply that the quantum states of the flag qubits can be an important element in the performance of quantum error-correcting codes in a  heavy-hexagon structure. In a future work, we will consider building a decoder using the correlation obtained from the result of measurement of the flag qubits in the tailored surface and XZZX codes.

\section*{Acknowledgements}
This work is supported by the Basic Science Research Program through the National Research Foundation of
Korea (NRF) funded by the Ministry of Education, Science and Technology (NRF2018R1D1A1B07049420 and
NRF2022R1F1A1064459) and Institute of Information and Communications Technology Planning and Evaluation
(IITP) grant funded by the Korean government (MSIT) (No. 2020001343, Artificial Intelligence Convergence Research Center (Hanyang University ERICA).

\section*{Data availability}
All data generated or used during the study appear in the submitted article.

\appendix

\section{Measurement circuit of Z stabilizer on heavy-hexagon structure}
To construct a surface code in a heavy-hexagon structure, the measurement circuits of the X and Z stabilizers should be built. Because the measurement circuits of the X stabilizer are discussed in the main text, in the Appendix, we explain 
the measurement circuits of the Z stabilizer.

\begin{figure*}[ht]
    \centering
    \includegraphics*[width=0.8\textwidth]{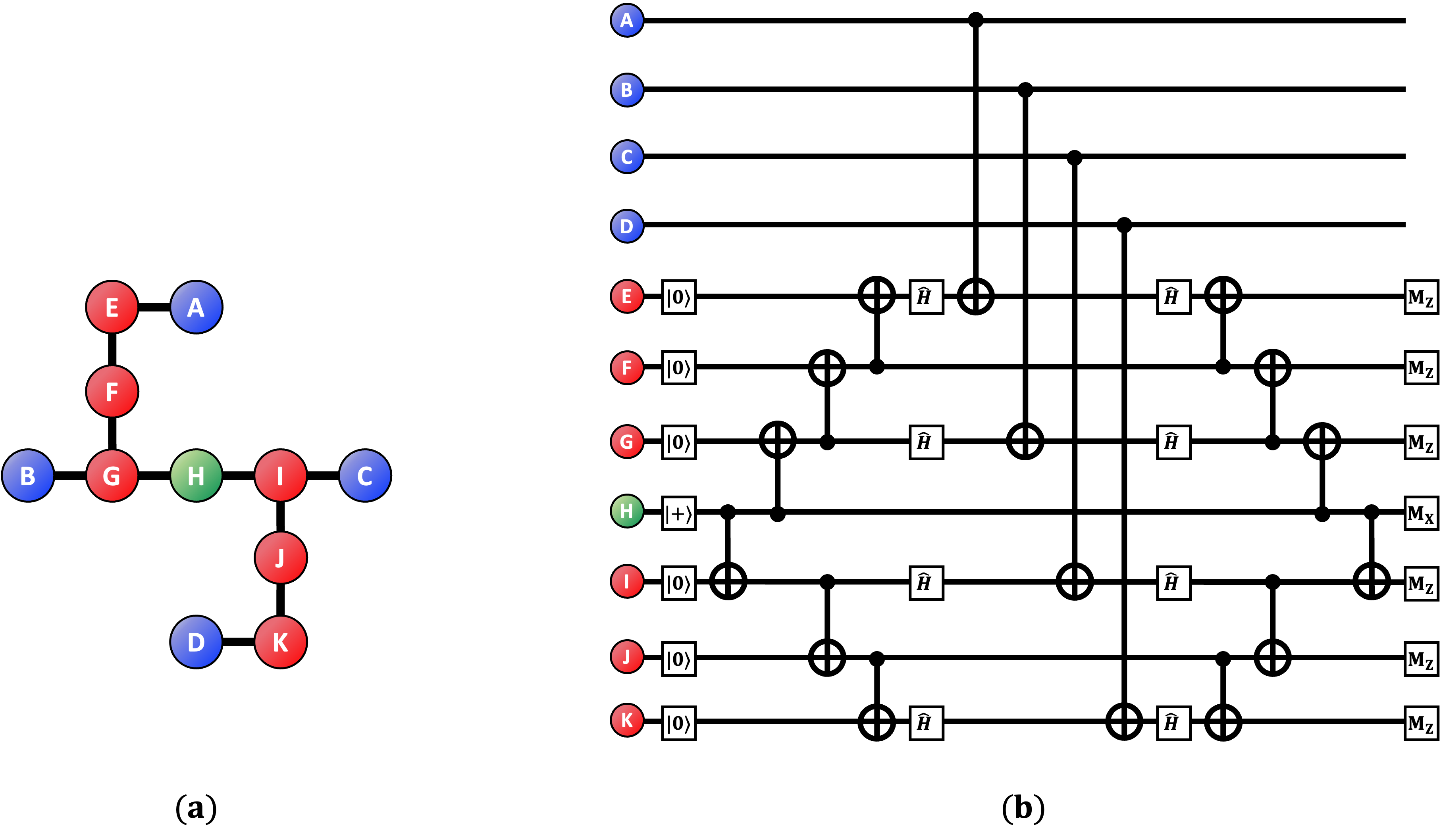}
    \caption{Z Stabilizer using six flag qubits in the heavy-hexagon structure. (a) Connection among qubits.  (b) Measurement circuit.
    The blue dots denote data qubits, and the red dots denote flag qubits. H is a syndrome qubit. For a measurement of the Z syndrome, the quantum states of the syndrome qubits(flag qubits) are prepared as $\ket{+}$($\ket{0}$).}
    \label{fig:11}
\end{figure*}

Fig.11.shows the assignment of qubits and gates to construct a measurement circuit for the Z syndrome in the heavy-hexagon structure. In the measurement circuit of the Z stabilizer, the X error is passed to the flag qubit through CNOT operator to evaluate the parity of the X error of the data qubit. 

The parity of the X error, which passed through the flag qubit to the syndrome qubit, is checked in the syndrome qubit. It should be noted that the quantum state of the flag qubits should be prepared as $\ket{0}$ and applied Hadamard operators, in order for the X error of the data qubit not to affect a flag qubit. The quantum state of the syndrome qubit is prepared as $\ket{+}$, to check the parity of the X error. In addition, in the measurement circuit of the Z stabilizer the CNOT operator is constructed using the flag qubit adjacent to a data qubit as the target qubit and the flag qubit adjacent to the syndrome qubit as the control qubit. The CNOT gate between a data qubit and a flag qubit is built using the data qubit as the control qubit and the flag qubit as the target qubit. 


\end{document}